\documentclass[journal=aesccq,manuscript=article,layout=twocolumn]{achemso}

\usepackage{chemformula} 
\usepackage[T1]{fontenc} 
\usepackage{graphicx}
\usepackage{tabularx}
\usepackage{amsmath}
\usepackage{amssymb}
\usepackage{natbib}

\author{G\'eraldine F\'eraud}
\email{geraldine.feraud@sorbonne-universite.fr}
\author{Mathieu Bertin}
\affiliation[Sorbonne]{Sorbonne Universit\'e, Observatoire de Paris, Universit\'e PSL, CNRS, LERMA, F-75005, Paris, France}
\author{Claire Romanzin}
\affiliation[LCP]{Laboratoire de Chimie Physique, CNRS, Universit\'e Paris-Sud, Universit\'e Paris-Saclay, 91405, Orsay, France}
\author{R\'emi Dupuy}
\author{Franck Le Petit}
\author{Evelyne Roueff}
\author{Laurent Philippe}
\author{Xavier Michaut}
\author{Pascal Jeseck}
\author{Jean-Hugues Fillion}
\affiliation[Sorbonne]{Sorbonne Universit\'e, Observatoire de Paris, Universit\'e PSL, CNRS, LERMA, F-75005, Paris, France}

\title[]
  {Vacuum ultraviolet photodesorption and photofragmentation of formaldehyde-containing ices}

\abbreviations{IR,TPD,FUV,VUV,ISM,PDRs,PPDs}

\begin{document}

\vskip12pt
\textbf{Accepted May 3, 2019 in ACS Earth and Space Chemistry, Special Issue: Complex Organic Molecules (COMs) in Star-Forming Regions}

\begin{tocentry}
    \includegraphics{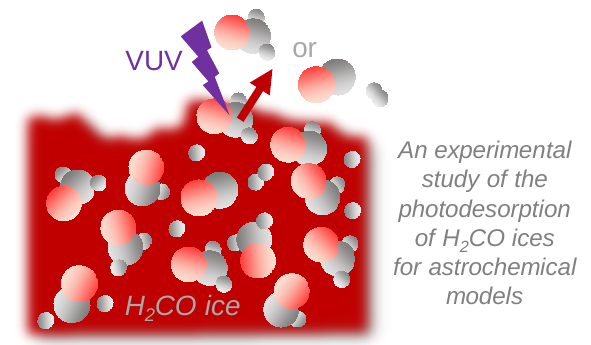}
\end{tocentry}


\begin{abstract}

Non-thermal desorption from icy grains containing H$_2$CO has been invoked to explain the observed H$_2$CO gas phase abundances in ProtoPlanetary Disks (PPDs) and Photon Dominated Regions (PDRs). Photodesorption is thought to play a key role, however no absolute measurement of the photodesorption from H$_2$CO ices were performed up to now, so that a default value is used in the current astrophysical models. As photodesorption yields differ from one molecule to the other, it is crucial to experimentally investigate photodesorption from H$_2$CO ices. 
 
We measured absolute wavelength-resolved photodesorption yields from pure H$_2$CO ices, H$_2$CO on top of a CO ice (H$_2$CO/CO), and H$_2$CO mixed with CO ice (H$_2$CO:CO) irradiated in the Vacuum UltraViolet (VUV) range (7-13.6~eV). Photodesorption from a pure H$_2$CO ice releases H$_2$CO in the gas phase, but also fragments, such as CO and H$_2$. 
Energy-resolved photodesorption spectra, coupled with InfraRed (IR) and Temperature Programmed Desorption (TPD) diagnostics, showed the important role played by photodissociation and allowed to discuss photodesorption mechanisms. For the release of H$_2$CO in the gas phase, they include Desorption Induced by Electronic Transitions (DIET), indirect DIET through CO-induced desorption of H$_2$CO and photochemical desorption.      

We found that H$_2$CO photodesorbs with an average efficiency of $\sim 4-10 \times 10^{-4}$ molecule/photon, in various astrophysical environments. H$_2$CO and CO photodesorption yields and photodesorption mechanisms, involving photofragmentation of H$_2$CO, can be implemented in astrochemical codes. The effects of photodesorption on gas/solid abundances of H$_2$CO and all linked species from CO to Complex Organic Molecules (COMs), and on the H$_2$CO snowline location, are now on the verge of being unravelled.

\end{abstract}

\section{Introduction}

Formaldehyde (H$_2$CO) is an organic molecule that was detected in the interstellar medium in 1969.\cite{snyder1969} It has been observed in several environments, such as Young Stellar Objects,\cite{roueff2006,araya2007} PDRs (e.g. Refs \citenum{leurini2010,guzman2011,cuadrado2017}), protostellar cores,\citep{maret2004} protoplanetary disks (e.g. Ref. \citenum{dutrey1997a}) and comets (e.g. Refs \citenum{snyder1989,meier1993}). It has been studied as a potential probe of planet formation in protoplanetary disks (PPDs),\citep{cleeves2015} and it is even used in cosmology as an extinction-free tracer of star formation across the epoch of Galaxy evolution.\citep{darling2012}
Its protonated form, H$_2$COH$^+$, detected for the first time in Sgr B2 and several hot cores,\cite{ohishi1996} has been recently observed in a cold prestellar core.\citep{bacmann2016a}
In the coma of several comets, its abundance has been found at the percent level relative to H$_2$O (e.g. Refs \citenum{snyder1989,mumma2011}), and its polymerized form, polyoxymethylene (POM), has also been detected.\citep{huebner1987,mitchell1987}  
In the solid phase, H$_2$CO is a likely identified species towards protostars, with the observation of a very weak IR absorption feature at 3.47 $\mu$m (C-H stretch),\citep{schutte1996} and of a stronger one at 5.83 $\mu$m (C-O stretch band), which is blended with a water band.\citep{ehrenfreund1997,keane2001,pontoppidan2004,boogert2015} The derived abundances are $\approx2-7\%$ relative to H$_2$O ice, which gives H$_2$CO/CH$_3$OH ice ratios of 0.09 to 0.51 in high-mass young stellar objects.\citep{ehrenfreund1997,dartois1999,keane2001,gibb2004,boogert2015} Towards the massive star W33, H$_2$CO has been detected both in the solid phase and in the gas phase, with a gas/ice ratio of 3~\%.\cite{roueff2006}

The origin of interstellar H$_2$CO appears to be manifold as it can be formed both on grains\citep{soma2018,guzman2011,guzman2018} and in the gas phase. This is different from CH$_3$OH, whose formation in the gas phase is considered inefficient,\cite{garrod2006,geppert2006,stoecklin2018} even if it could be at play in stellar outflows or in shocked circumstellar regions.\citep{dartois1999} 
On grains, H$_2$CO formation depends on the competition between H addition and desorption (e.g. Refs \citenum{minissale2016a,morisset2019}). The location of H$_2$CO in the icy mantles of the interstellar grains changes with many parameters such as time, temperature of the grain, n$_H$/n$_{CO}$ gas phase abundance, grain porosity, \citep{cuppen2009,taquet2012} and as a consequence it evolves with the processing of the interstellar ice. It could be in the upper layers of the ice, together with CO, CO$_2$ and CH$_3$OH, or it could also be buried deeper in the ice, possibly with H$_2$O (Refs \citenum{boogert2015,cuppen2009} and references therein).

Laboratory studies have given a wealth of information regarding the formation of solid H$_2$CO from various ices, using different processes: 
(i) by the hydrogenation of CO ices \citep{watanabe2002,fuchs2009,pirim2011}
which has been recently revisited  
\citep{minissale2016a} 
(ii) by processing of CH$_3$OH ices either by ion irradiation,\citep{deBarros2011} by UV photons \citep{oberg2009b} or by soft X-ray photons \citep{ciaravella2010}
(iii) by processing of H$_2$O:CO ice mixtures, either by UV photons \citep{schutte1996}, by protons \citep{hudson1999} or by electrons.\citep{yamamoto2004} 
Some experiments also unravelled the evolution of solid phase H$_2$CO under specific conditions, for example by reactions with oxygen atoms (leading to CO$_2$),\citep{minissale2015} or by VUV irradiation.\cite{gerakines1996,butscher2016} VUV-lamp irradiation of pure H$_2$CO ices or of H$_2$CO-containing ices showed the formation of many products within the ice, showing the richness and the complexity of photochemistry in the solid phase,\cite{gerakines1996,butscher2016} even for a small organic molecule such as H$_2$CO.

It has been proposed that hydrogenation of H$_2$CO forms CH$_3$OH, so that H$_2$CO and CH$_3$OH are both of interest as precursors of larger organic molecules, the so-called Complex Organic Molecules (COMs) (e.g. Ref. \citenum{herbst2009}). Understanding the formation of COMs is the objective of several experiments
 \citep{theule2013,butscher2016,chuang2017} and of several models, including grain surface reactions or possibly gas-phase reactions,\cite{woods2012,vasyunin2013,balucani2015,ruaud2015,kalvans2015a}
but the recent observation of COMs in the gas phase in  pre-stellar cores, which are cold environments, was quite puzzling.\cite{bacmann2012,vastel2014,jimenez-serra2016} Their presence could be explained by non-thermal desorption, however the nature of the desorption mechanisms still has to be unravelled, with a joint experimental and modelling effort.

Non-thermal desorption from icy grain surfaces includes chemical desorption (also called reactive desorption, that is a chemical reaction at the surface of the ice forming a product that could desorb if exothermicity is sufficient)\cite{minissale2016b,minissale2016a}, desorption by cosmic rays,\cite{dartois2015} desorption by shock sputtering,\cite{gusdorf2008} photodesorption by UV (6-13.6~eV) primary photons or by secondary photons from cosmic rays and photochemical desorption (where photoproducts at the surface of the ice react and desorb).\cite{fillion2014,martin-domenech2016} We can also add photodesorption by soft X-ray photons which has recently been experimentally explored and suggested to be important in PPDs.\cite{dupuy2018} UV photodesorption from ices has been proposed to play a key role in the gas-to-ice balance in various environments such as PDR, PPDs and protostellar envelopes (eg Ref.~\citenum{hollenbach2009}), and also in the determination of snowlines.\cite{ligterink2018}

If one considers that H$_2$CO is a precursor of COMs, it is of prime importance to understand its photochemistry and the gas-grain exchanges it is involved in, as it could impact the formation and evolution of COMs. Even for a small molecule like H$_2$CO, gas-grain interactions are complex, and the gas-to-ice ratio results from the competition between thermal desorption, non-thermal desorption and freeze-out/processing of dust grains. It is crucial to understand in particular to which extent H$_2$CO can photodesorb as intact or dissociated in photofragments.

UV photodesorption of H$_2$CO has been called upon to explain molecular abundances and spatial distribution in PDR\citep{guzman2011,guzman2013} and in PPDs.\citep{loomis2015,carney2017,oberg2017} 
There is a large number of astrochemical models that include H$_2$CO photodesorption, \citep{guzman2011,guzman2013,walsh2014,agundez2018,esplugues2016,esplugues2017,koumpia2017} but an arbitrary H$_2$CO photodesorption yield of 10$^{-3}$ molecules/photon is often considered, due to the absence of experimental data. However knowledge of photodesorption yields is crucial, as its variations, for the formaldehyde molecule, could increase gas phase abundances by up to several orders of magnitude in some environments, 
 \citep{guzman2011} together with changes in the solid-phase chemistry.\citep{esplugues2016,esplugues2017}
It is clear that photodesorption yields from pure ices do vary on orders of magnitude from one molecule to the other, for example it is $\sim 10^{-2}$ molecules/photon for CO \cite{fayolle2011} and $\sim 10^{-5}$ molecules/photon for CH$_3$OH. \cite{bertin2016,cruz-diaz2016} Photodesorption has been well studied for diatomics (CO,\cite{fayolle2011,dupuy2017a} N$_2$,\cite{fayolle2013a,bertin2013} NO),\cite{dupuy2017} but is much less known for larger molecules, and especially for organic molecules. Besides, photodesorption of a given molecule also varies with ice composition.

Whereas several experimental works focus on the photochemistry of H$_2$CO-containing ices, which is bulk sensitive, despite a clear need none studied the interaction between H$_2$CO ices and gas, which essentially involves the surface of the ice.  
 To our knowledge, measurements of the photodesorption of H$_2$CO in the gas phase were performed from ices containing H$_2$CO as a photoproduct of the irradiation of pure methanol ices,\citep{bertin2016,cruz-diaz2016} ethanol ices, or H$_2$O:CH$_4$ ice mixtures,\citep{martin-domenech2016} but never from ices grown from H$_2$CO, so that the amount of H$_2$CO in these ices is not well controlled.

We present wavelength-resolved and quantified data on the photodesorption of H$_2$CO and of CO and H$_2$ fragments from model ices : pure H$_2$CO ice, H$_2$CO on top of a CO ice (H$_2$CO/CO), and H$_2$CO mixed with CO ice (H$_2$CO:CO). Absolute photodesorption yields for each desorbing species, H$_2$CO, CO and H$_2$, are derived for each ice. Based on the results obtained from the pure H$_2$CO ice and the ones containing H$_2$CO and CO, we explore possible photodesorption mechanisms. Average photodesorption yields and branching ratios are deduced in various astrophysical environments (InterStellar Radiation Field ISRF, PDR at different extinctions, dense cores and PPDs). We discuss the possible effects of the  photodesorption of H$_2$CO and of CO fragments in dense cores, PDR and PPDs.

\section{Experiment}

The SPICES (Surface Processes \& ICES) set-up, described in \citet{doronin2015}, was used for these experiments. It consists of an ultra-high vacuum (UHV) chamber with a base pressure of typically $10^{-10}$ mbar, within which a polycrystalline gold surface is mounted on a rotatable cold head that can be cooled down to $\sim$~10~K using a closed cycle helium cryostat. Several diagnostics are possible in this set-up : detection of photodesorbing neutral molecules in the gas phase through mass-spectrometry with a quadrupole mass spectrometer (QMS, from Balzers), IR spectroscopy of the ice with Reflection Absorption InfraRed Spectroscopy (RAIRS), and temperature programmed desorption (TPD) of the ice with the same QMS as used for photodesorption measurements.\citep{doronin2015}

\begin{figure}
    \includegraphics[width=8.26cm]{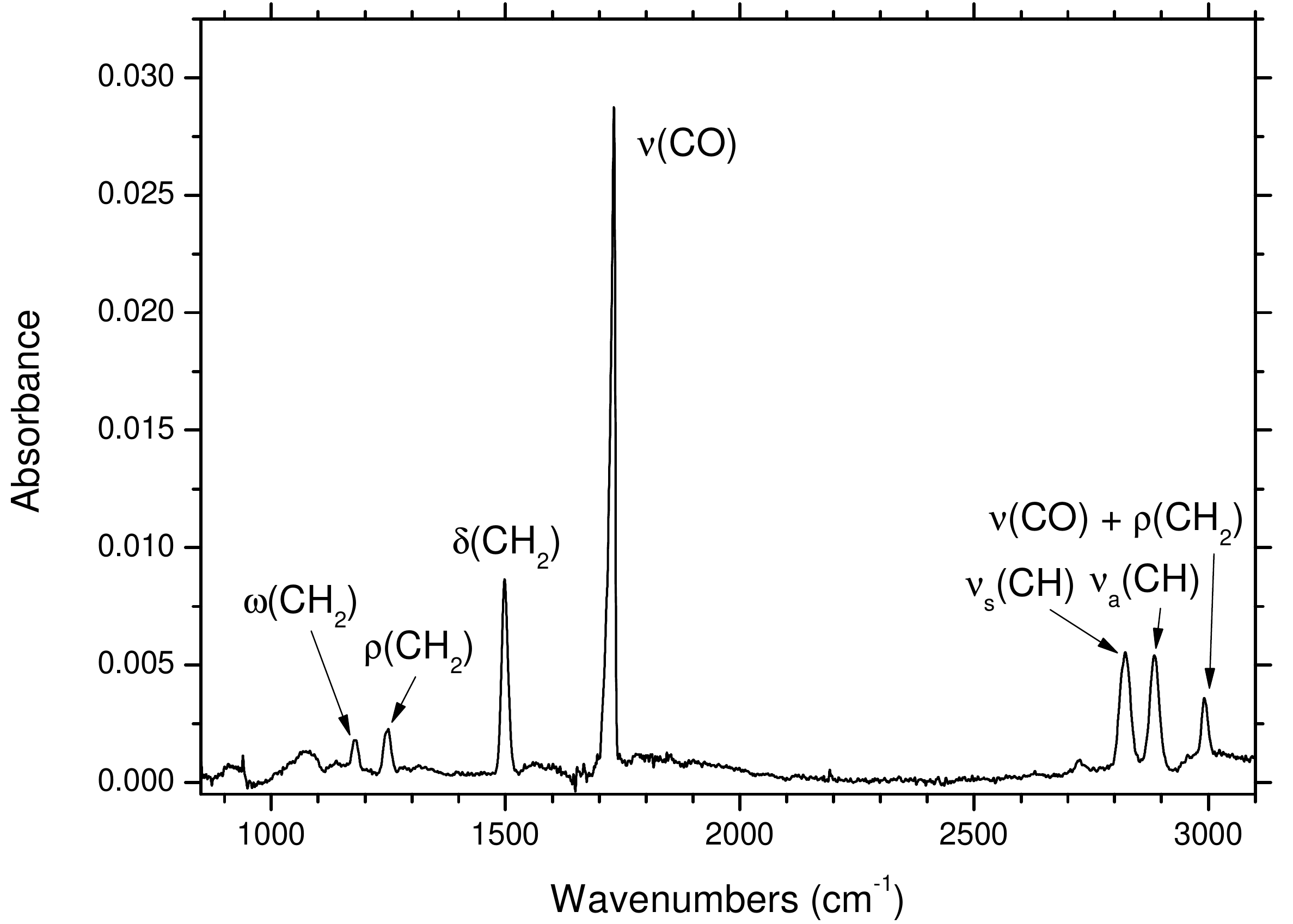}
    \caption{RAIRS spectrum of a 15~ML thick H$_2$CO ice (baseline subtracted). Vibrational modes are assigned according to \citet{truong1993}. $\nu(CH)$ are stretching vibrational modes (either symetric ($s$) or asymetric ($a$)). $\omega (CH_2)$, $\rho (CH_2)$ and $\delta (CH_2)$ are wagging, rocking and scissoring modes of the CH$_2$ group respectively.  
    }
\label{fig_IR_attribution}
\end{figure}

Ices are dosed by exposing the cold surface (10~K) to a partial pressure of gas using a tube positioned a few millimeters in front of the surface, allowing rapid growth without increasing the chamber pressure to more than a few 10$^{-9}$ mBar. In this set of experiments, different ices were grown : pure H$_2$CO ices, layered ices where H$_2$CO is deposited on a CO ice (H$_2$CO/CO)\footnote{We have grown three different ices, in the following order : (1) 0.4 ML of H$_2$CO on CO, (2) 1.1 ML of H$_2$CO on CO, and (3) 2.6 ML of H$_2$CO on CO; to obtain the second ice, we added 0.7 ML of H$_2$CO on the first one, and to obtain the third one, we added 1.5 ML to the second one} and mixed H$_2$CO:CO ices, where H$_2$CO and CO gases are mixed prior to deposition. Ice thicknesses are based on those measured from CO (Air liquide, >99.9\% purity) ices, that are controlled with a precision better than 1 monolayer (ML) via a calibration using TPD, as detailed in \citet{doronin2015}. 
Condensing a pure H$_2$CO ice is not easy. Several tests have been performed in order to obtain the purest possible ice, and the following was performed (note that other techniques exist). First, solid paraformaldehyde, a (OCH$_2$)$_n$ multimer of formaldehyde, was pumped with primary pumps at room temperature for several hours. Then, its temperature was gradually increased up to 60~$^{\circ}$C with a water bath. Additional turbomolecular pumping is performed before the introduction of gas phase H$_2$CO to ensure the elimination of remaining water. This purified H$_2$CO in the gas phase was then released in the injection circuit and condensed on the cold substrate. The purity was checked with mass spectrometry and with IR spectroscopy of the pristine ice, with the RAIRS technique (a gold surface has been chosen for its good reflectance properties in the IR, so that RAIRS spectroscopy could be performed). A typical IR spectrum is shown in Fig.\ref{fig_IR_attribution}. It confirms that H$_2$CO is the main component of the ice, with no H$_2$O or polymer contribution. Different H$_2$CO vibrational bands are labelled in the figure.\cite{bouilloud2015} For each new ice deposition, fresh  H$_2$CO in the gas phase was expanded in the injection circuit.

The chamber was coupled to the undulator-based DESIRS beamline \citep{nahon2012} at the SOLEIL synchrotron facility, which provides monochromatic, tunable VUV light for irradiation of our ice samples. The coupling is window-free to prevent cut-off of the higher energy photons. The size of the VUV beam on the gold surface is $\sim0.7$ cm$^2$.\cite{fayolle2011} To acquire photodesorption spectra, the narrow bandwith ($\sim25$ meV) output of a grating monochromator is continuously scanned between 7 and 13.6 eV. Higher harmonics of the undulator are suppressed using a Kr gas filter. The experimental procedure is the following : we deposit an ice, record an IR spectrum to check its purity, irradiate it with VUV from 7 to 13.6 eV and record the photodesorption signal as a function of energy. 
The photodesorption of molecules in the gas phase following VUV irradiation of the ices is monitored by means of the QMS. Each 25 meV photon energy step lasts about 5 s, which is sufficiently higher than the dwell time of the QMS (0.5 s). 
Typical photon fluxes, as measured with a calibrated AXUV photodiode, depend on the photon energy and vary between $1.3 \times 10^{13}$ photons~cm$^{-2}$~s$^{-1}$ at 7 eV and $5 \times 10^{12}$ photons~cm$^{-2}$~s$^{-1}$ at 10.5 eV. A typical energy scan from 7 to 13.6 eV thus lasts around 20 minutes, which corresponds to a fluence of $\sim$ 10$^{16}$ photons~cm$^{-2}$. In the following, a 'fresh' ice refers to an ice that has just been deposited. We also performed several VUV irradiations on the same ice, further qualified as 'aged'. IR spectra after VUV irradiations are recorded and at the end, the ice and its photoproducts are released in the gas phase through TPD.

The conversion from the QMS signal to the absolute photodesorption efficiency, in molecule per incident photon, has been described in detail in \citet{dupuy2017a}. It is based on the knowledge of the absolute VUV photon flux, the apparatus function of the QMS, the 70~eV electron-impact ionization cross sections and comparative measurements to well-known photodesorption yields measured in the same experimental conditions. Due to an uncertainty in the photon flux measurement between 9.5 and 10.5~eV, the shape of the photodesorption spectrum has to be taken with caution in this region (see Supporting Information). In the electron-impact ionization process, H$_2$CO gives H$_2$CO$^+$, but also HCO$^+$ and CO$^+$. So any signal measured on the HCO$^+$ and on the CO$^+$ channels is corrected from this cracking. For all the detected species, electron-impact ionization cross-sections\cite{vacher2009,tian1998a,straub1996} were used to obtain absolute photodesorption yields. We should note that in the particular case of H$_2$, the apparatus function of the QMS is not known well and the partial electron-impact ionization cross-section of H$_2^+$ from H$_2$CO$^+$ seems unknown, so that photodesorption yields of H$_2$ from H$_2$CO are indicative values, from which not too many conclusions can be drawn.

\section{Results}

\subsection{Pure H$_2$CO ices}
\subsubsection{Photodesorption spectra of fresh H$_2$CO ices}
\label{section_fresh}

\begin{figure}
   \includegraphics[width=8.26cm]{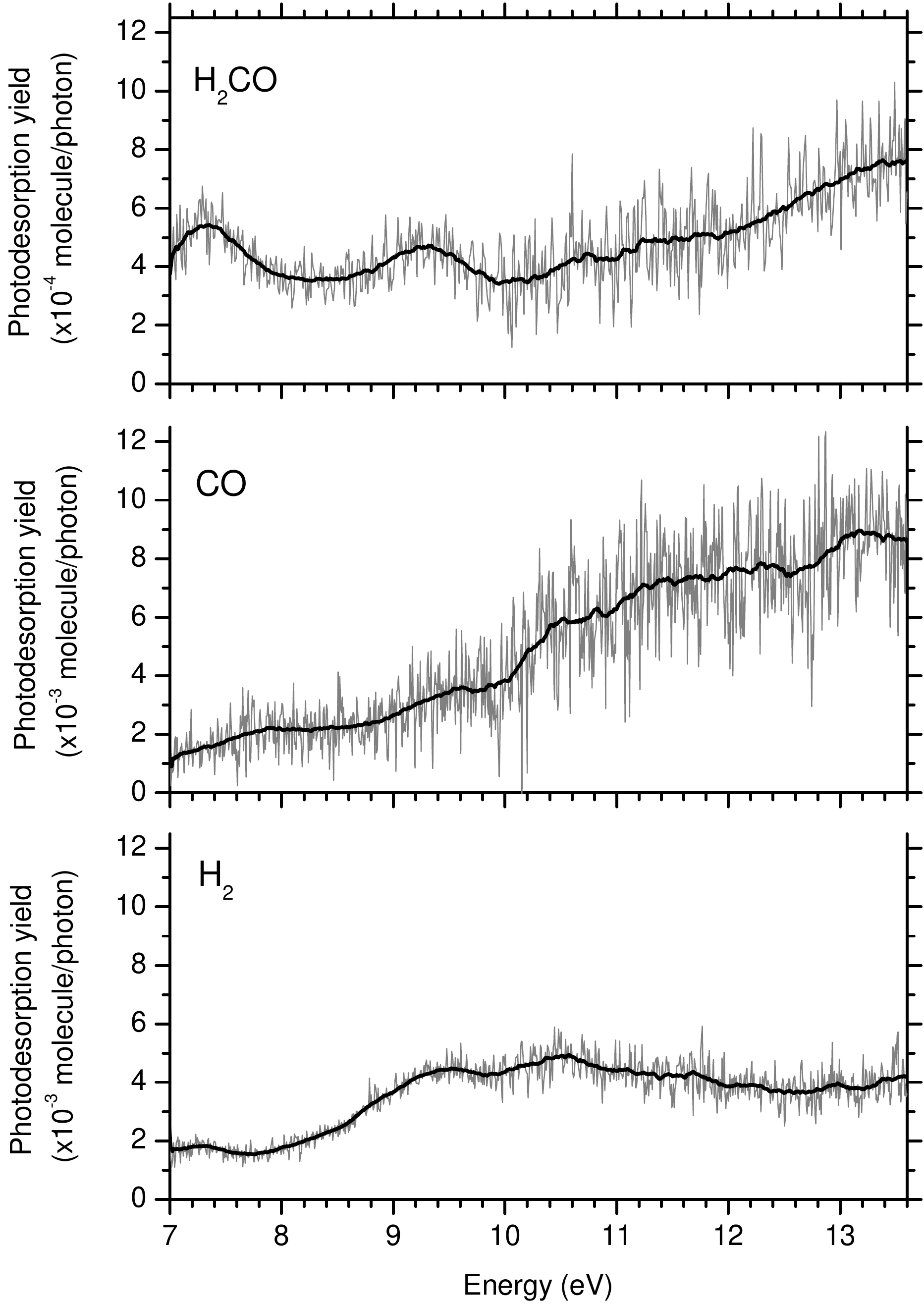}
    \caption{ Absolute photodesorption spectra of a 15 ML thick fresh H$_2$CO ice on gold at 10~K between 7 and 13.6~eV, recorded on  the H$_2$CO mass (m/z 30, \textit{upper panel}) and the CO mass (m/z 28, \textit{middle panel}). The photodesorption spectrum recorded on the H$_2$ mass (m/z 2, \textit{lower panel}) gives indicative yields (see Experiment). Note that the vertical scale on the upper panel is ten times lower than that in the other two panels. Spectra are averaged on three energy scans. Smoothed data (adjacent-averaging on 40 points) are represented by a bold black line.
    }
     
    \label{fig_pure_H2CO}
\end{figure}

\begin{figure}
    \includegraphics[width=8.26cm]{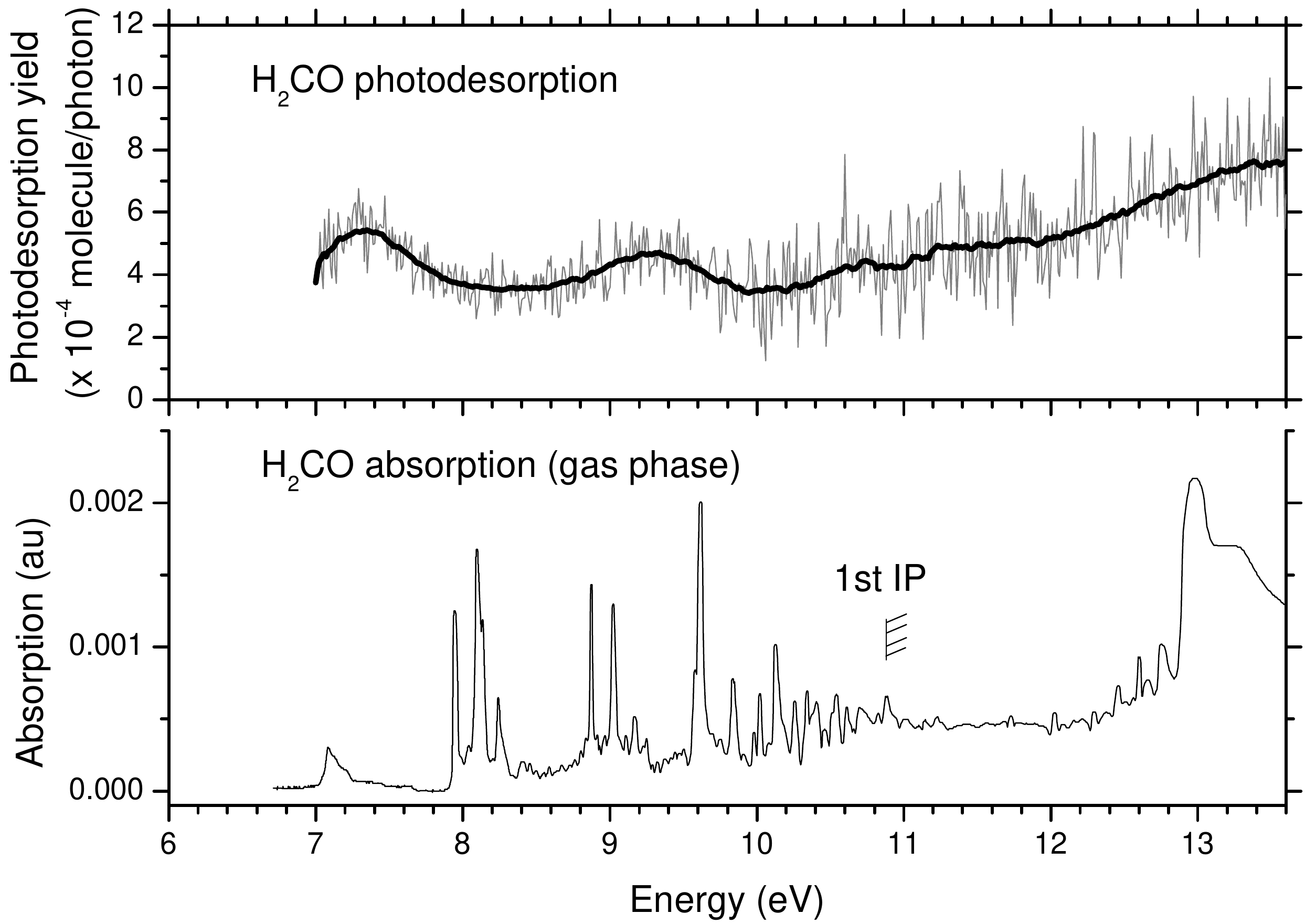}
    \caption{ Comparison between the H$_2$CO photodesorption spectra of a 15 ML thick H$_2$CO ice (upper panel; same as Fig.~\ref{fig_pure_H2CO}) and H$_2$CO gas phase absorption spectrum from \citet{gentieu1970} (lower panel). The first ionization potential (1st IP) of gaseous H$_2$CO is indicated with a vertical line (10.88~eV \citep{kimura1981}).}
    
    \label{fig_H2CO_gentieu}
\end{figure}

First of all, we recorded photodesorption spectra of an as-deposited H$_2$CO ice (Fig.\ref{fig_pure_H2CO}). Several molecules were detected in the gas phase following VUV irradiation : the parent molecule, H$_2$CO, and fragments, CO and H$_2$. No signal of HCO photodesorption from H$_2$CO ices was detected. Indeed, whereas a HCO signal was measured, once corrected from the cracking pattern of H$_2$CO by the electron impact ionization, the signal on the HCO channel is dominated by noise. We estimated an upper limit of HCO photodesorption at $\sim 5 \times 10^{-4}$ molecule/photon, assuming the ionization cross section of HCO to be the same as that of H$_2$CO. The fact that H$_2$CO molecules are photodesorbing is noticeable. The photodesorption yield of H$_2$CO is around $5 \times 10^{-4}$ molecule/photon. Besides, the detection of CO and H$_2$ fragments is an important result and shows that VUV photons photodissociate a part of H$_2$CO ice in CO and in H$_2$, and that the CO fragments can desorb efficiently. The CO photodesorption intensities, from $10^{-3}$ to $10^{-2}$ molecule/photon, are indeed higher than those measured on the H$_2$CO channel.

The H$_2$CO photodesorption spectrum presents two broad structures around 7.2 and 9.2 eV, and a continuous increase above 10~eV. Comparing the parent photodesorption spectrum with solid phase absorption usually gives useful insights for the interpretation of the photodesorption spectrum, as the absorption is the first step eventually leading to photodesorption. Due to the lack of VUV absorption spectrum of solid H$_2$CO, the H$_2$CO photodesorption spectrum can only be discussed in the light of the H$_2$CO gas phase absorption. Figure~\ref{fig_H2CO_gentieu} presents the VUV absorption spectrum of H$_2$CO in the gas phase obtained with the monochromatized output of Hydrogen or Helium lamps (resolution 1\AA) from \citet{gentieu1970}, and the reproduction of H$_2$CO photodesorption spectrum from Fig. \ref{fig_pure_H2CO} for comparison. 
In the gas phase, there is a first electronic state (n,$\pi\star$) that shows a very weak UV absorption between 240 and 360~nm (3.4-5.2~eV, not shown). The highly-structured H$_2$CO spectrum between 115 and 160 nm (7.7-10.8~eV) consists of predissociated Rydberg series (eg Ref.~ \citenum{brint1985}). These series are formed via excitation from the (2b$_2$) nonbonding orbital, and are accompanied by little vibrational excitation of the ionic core. On the contrary, Rydberg states in the 12-14~eV region enhance the production of vibrationally excited ions i.e. there could be autoionization from these Rydberg states,\citep{praet1968,brint1985,holland1994} and a strong and large resonance is seen in the gas phase spectrum (Fig.~\ref{fig_H2CO_gentieu}). Gas phase studies in the VUV show that when exciting H$_2$CO at 13~eV, the production of ions is favored relatively to that of neutrals and H$_2$CO$^+$ is the major ion detected, HCO$^+$ being much smaller.\citep{tanaka2017}

The fact that the electronic states of H$_2$CO in the gas phase are predissociative, the large width of the features observed in the photodesorption spectrum (Figure \ref{fig_pure_H2CO}), and the observation of photodesorbing fragments point to the dissociative nature of the electronic states of solid H$_2$CO.
Another finding is that photodesorption yields lie in the $4-8 \times 10^{-4}$ molecule/photon in the 7-13.6~eV range, and increase slightly and continuuously above 10~eV, which is expected to correspond more or less to the ionization energy (the ionization energy of solid H$_2$CO is not known, but is expected about 1~eV below the gas phase value of 10.88 eV). \cite{kimura1981}   
Besides, it seems that no sign of an autoionized state (which is at around 13 eV in the gas phase) is observed in the photodesorption spectrum, so that autoionization is not correlated to the photodesorption of neutral molecules. Possible causes include a very short lifetime in the solid phase and/or an electronic transition that does not lead to desorption of neutral H$_2$CO or of CO fragments. Regarding the possible photodesorption of ions, one has to keep in mind that ions in the solid phase are more bound than neutrals for polar ices, and that the electron takes the major part of the energy in the ionization process, so that the photodesorption of ions requires more energy than the ionization energy.\citep{philippe1997}
Yet, desorption mechanisms when energies are higher than the ionization energy are poorly known. The detection of photodesorbing ions is not optimized for this version of the SPICES set-up, and no attempt to detect ions was performed.

CO and H$_2$ photodesorption spectra from fresh H$_2$CO ice show a different shape than the H$_2$CO photodesorption spectrum (Fig.~\ref{fig_pure_H2CO}). The H$_2$ spectrum presents an increase up to 9.5 eV, and then it remains at a constant yield of $4 \times 10^{-3}$ molecule/photon. However, there is a strong uncertainty on the absolute intensity and on the shape of H$_2$ photodesorption spectrum. This is due to the correction of the H$_2$ photodesorption measurement by a strong baseline coming from the residual vacuum.
Finally, the CO photodesorption spectrum from pure H$_2$CO ices does not look like the photodesorption spectrum from pure CO ices. Indeed, no structures are seen, but only a strong increase with the photon energy, from $1 \times 10^{-3}$ molecule/photon at 7 eV to $9 \times 10^{-3}$ molecule/photon at 13.6 eV (Fig. \ref{fig_pure_H2CO}) whereas the CO photodesorption spectrum from pure CO ices is null at 7~eV and between 9.5 and 10.5 eV, and show characteristic peaks between 8 and 9 eV (it is reproduced from \citet{dupuy2017a} in Figure \ref{fig_H2CO_CO}). CO desorption here thus proceeds through H$_2$CO excitation. Possible photodesorption mechanisms of H$_2$CO and of CO will be further developped in the discussion.

\subsubsection{Ice modifications with VUV fluence (ice ageing): bulk and surface diagnostics through Infrared spectoscopy, TPD, and photodesorption spectra}
\label{section_IR_TPD}

In this subsection, the results obtained when irradiating an ice that was already irradiated (ie an 'aged' ice) are described. The study of ice modifications could be performed both by probing the bulk of the ice, with IR spectroscopy and TPD, and the surface of the ice through repeated photodesorption spectra on the same ice.

\begin{figure*}
    \includegraphics[width=\textwidth]{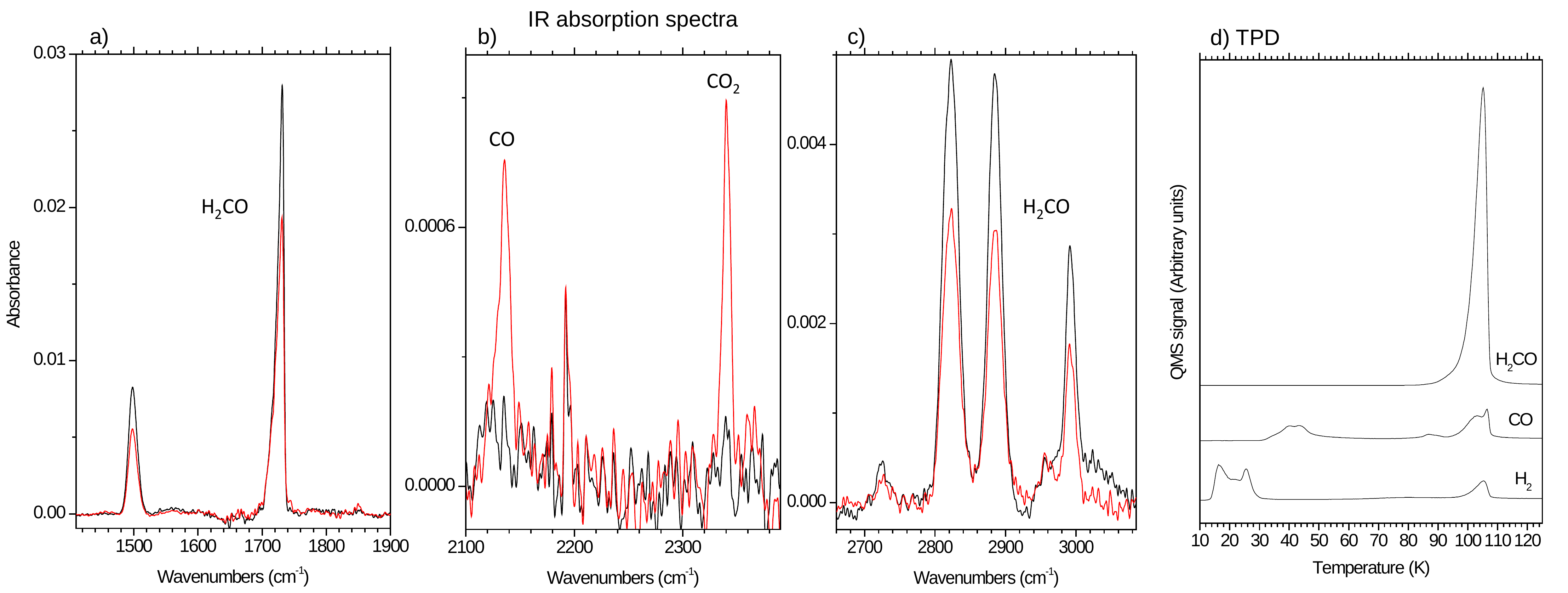}
    \caption{Diagnostics of the ice modifications within VUV irradiation of a H$_2$CO ice which received a fluence of $3 \times 10^{16}$ photons cm$^{-2}$. (a)-(c) RAIRS spectra of an H$_2$CO ice before VUV irradiation (black line) and after irradiation (red line). A baseline has been subtracted. Three different spectral regions are represented. In the 1400--1900 cm$^{-1}$ region (a) and 2700--3100 cm$^{-1}$ region (c), decreasing of H$_2$CO bands is seen. In the 2100--2400 cm$^{-1}$ region (b), there is formation of several photoproducts : CO (2138 cm$^{-1}$) and CO$_2$ (2341 cm$^{-1}$). (d) Temperature programmed desorption (TPD) after irradiation, measured on H$_2$CO, CO and H$_2$ masses. The signal on the CO mass is corrected from the signal coming from the cracking of H$_2$CO, whereas that on the H$_2$ mass is not, and no quantification from the latter is performed. The desorption peak of H$_2$CO is at 105~K.      
    }
    
    \label{fig_IR_irradiation}
\end{figure*}

IR spectra and Temperature Programmed Desorption give interesting results on the formation of photoproducts in the bulk of the ice. The comparison between IR spectra of a pristine ice and of an ice which received $3 \times 10^{16}$ photons cm$^{-2}$ is presented in Figure \ref{fig_IR_irradiation}. The diminution of H$_2$CO is clearly seen in the 1400--1900 cm$^{-1}$ and in the 2800-3100 cm$^{-1}$ region. As the IR light illuminates the whole ice area (1~cm$^2$), but the VUV irradiates only 70\%, this has to be taken into account in the analysis of IR spectra. We can estimate that approximately 2.4~ML of H$_2$CO was removed assuming a constant oscillator strength with the fluence. Given the initial thickness of 15~ML, this  gives  $\sim$ 20 \% of removal of H$_2$CO at this fluence.
The formation of CO and CO$_2$ in the ice was also evidenced with IR spectroscopy, by the appearance of characteristic bands at 2138 cm$^{-1}$ and 2341 cm$^{-1}$, respectively (Figure \ref{fig_IR_irradiation}). The detection of HCO fragments in the ice with IR spectroscopy is not conclusive in our experiments (signal to noise ratio too low).

Figure \ref{fig_IR_irradiation} also shows the TPD of the same irradiated ice, which received a total fluence of $3 \times 10^{16}$ photons cm$^{-2}$. Together with the thermal desorption of H$_2$CO, desorption of the photoproducts H$_2$ and CO was observed. The signal measured on CO is corrected from the cracking of H$_2$CO. H$_2$CO desorbs around 105~K, and the co-desorption of CO and H$_2$ together with H$_2$CO at this temperature is also noticed. CO also desorbs at lower temperature, at $\sim$ 40 K, which is larger than its desorption temperature from pure CO ices. That could be due to its diffusion in the H$_2$CO ice, or its presence in H$_2$CO pores. TPD is very useful as a quantitative technique but it may be artificially altered by thermal activation of chemical reactions. The amount of CO fragments in the bulk of the irradiated H$_2$CO ice is estimated at $\sim 20\%$ in abundance from TPD measurements. The absence of higher masses rules out POM presence.

In the end, the IR spectra and TPD give compatible amounts for the photodissociation of H$_2$CO in CO, $\sim 20\%$. The carbon and oxygen budget is balanced in the irradiation process. H$_2$CO disappearance in the bulk is due to the photodissociation of H$_2$CO, which mainly leads to CO formation (see Discussion). The amount of CO and H$_2$CO ejected in the gas phase is negligeable as compared with the amount of material within the bulk, as it will be shown below.

\begin{figure}
    \includegraphics[width=8.26cm]{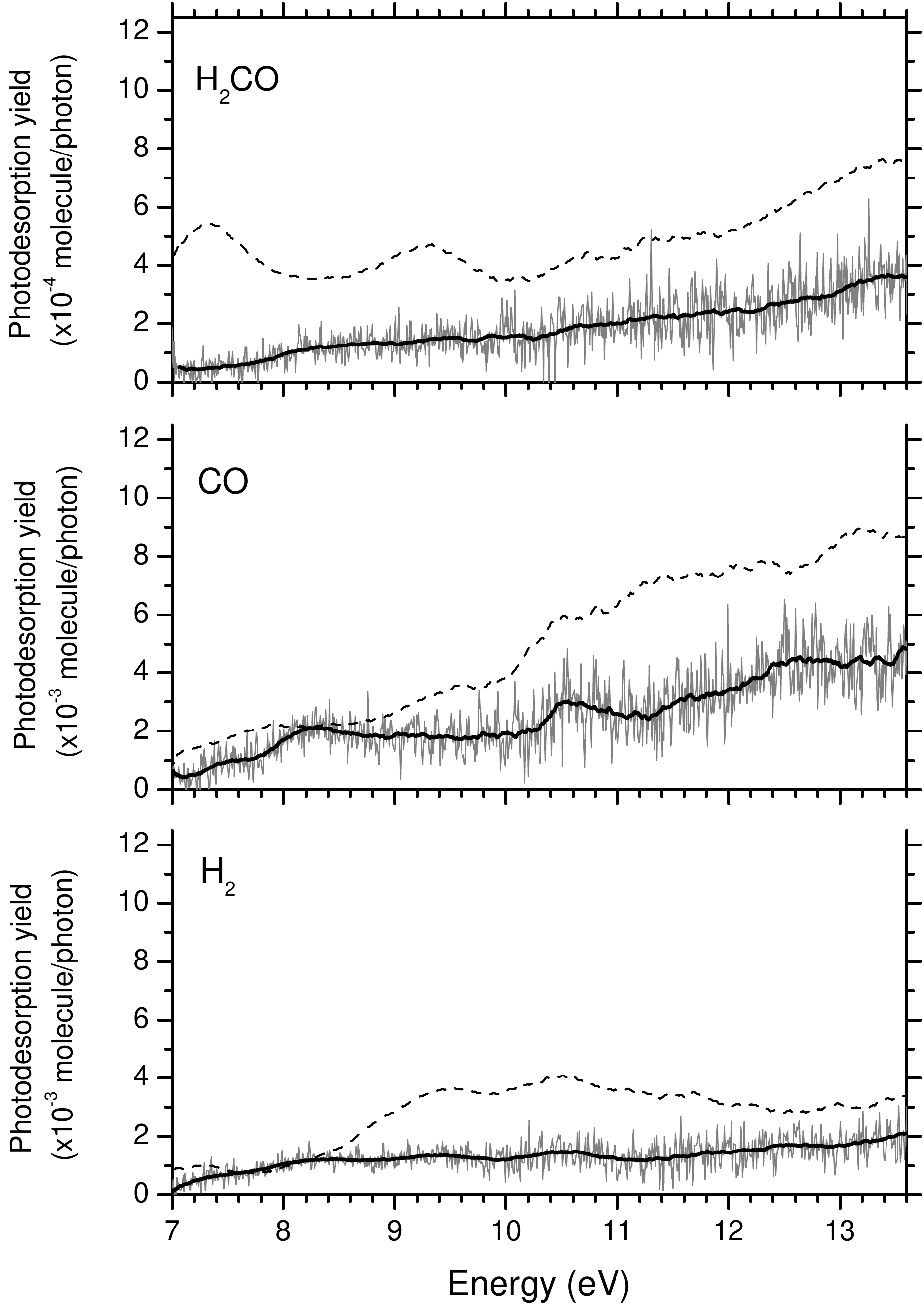}
    \caption{ Same as Fig.~2, but for an ice pre-irradiated (fluence from $2 \times 10^{16}$ photons cm$^{-2}$ at 7~eV to $3 \times 10^{16}$ photons cm$^{-2}$ at 13.6~eV.) Smoothed data from the 'fresh' ice of Fig.~2 are reproduced in dashed lines. 
    }
   
    \label{fig_pure_H2CO_aged}
\end{figure}

As mentioned earlier, ice modifications due to VUV irradiation are noticed not only in bulk diagnostics, but also in the signatures of the surface processes, through photodesorption. Figure~\ref{fig_pure_H2CO_aged} presents photodesorption spectra for an ice which was already irradiated (at a fluence of $2 \times 10^{16}$ photons cm$^{-2}$) so that while recording the spectrum, the fluence varied between 2 and 3 $\times 10^{16}$ photons cm$^{-2}$. We qualify such an ice as 'aged', contrary to 'fresh' ices of Fig.~\ref{fig_pure_H2CO}. From this figure, it is clear that there are changes in the intensity and shape of the spectra, showing that the VUV irradiation photoprocessed the ice. The VUV flux used here is thus high enough to process the ice within one energy scan.
More precisely, it should be noted that photodesorption spectra at the intermediary fluence (1 to 2 $\times 10^{16}$ photons cm$^{-2}$ i.e. just between those of Figure \ref{fig_pure_H2CO} and \ref{fig_pure_H2CO_aged}) look very much like the spectrum of the Figure \ref{fig_pure_H2CO_aged} (see Supporting Information). Here a steady state may be reached certainly during the second energy scan. 
The fact that a steady state was not reached during the first energy scan in the 'fresh' ice shows that the shape of the H$_2$CO spectrum during the first energy scan is certainly distorted, especially at the end of the scan. However, due to the weak H$_2$CO signal, it was not possible to perform measurements at lower flux.

The spectral shape of CO changes as a function of fluence (Figures \ref{fig_pure_H2CO} and \ref{fig_pure_H2CO_aged}). The bump in the 8--9 eV region (Figure \ref{fig_pure_H2CO_aged}) could be the signature of accumulated solid CO electronic A-X transition. This is consistent with the well-known photodesorption signatures of CO from CO ices (see for example Ref.~\citenum{fayolle2011} and Figure \ref{fig_H2CO_CO}) and with the fact that CO is present in the ice following VUV dissociation of H$_2$CO, as shown by IR and TPD measurements. This is another signature of the photoprocessing of the ice.  
Interestingly, whereas for fresh ices (Figure \ref{fig_pure_H2CO}), H$_2$CO, CO and H$_2$ photodesorption spectra look different, for aged ices their spectral shape looks much alike (Figure \ref{fig_pure_H2CO_aged}).

From the VUV irradiation dose and the average photodesorption yield, it is possible to estimate the amount of desorbed species during one energy scan: $4 \times 10^{-3} $ ML of H$_2$CO desorbed, with $4 \times 10^{-2} $ ML of CO and with $3 \times 10^{-2} $ ML of H$_2$. This corresponds to a total of only 7\% of a ML desorbing during one energy scan, which is very small compared to the ice thickness, accounting for only 0.5\%. It is also very small compared with the disappearance of H$_2$CO in the bulk, estimated at 20\% from IR spectra. The amount of photodesorbing molecules is thus very small compared with the amount of photoprocessed ones; the overall ice thickness can thus be considered as constant. This implies that the decrease of the photodesorption signal between a fresh and a photoprocessed ice (Figures \ref{fig_pure_H2CO} and \ref{fig_pure_H2CO_aged}) cannot be related to the evolution of the ice thickness. The loss of solid H$_2$CO through photodissociation followed by chemistry, together with a possible modification of the ice surface may thus be responsible for the decrease of the H$_2$CO photodesorption signal.


\subsection{H$_2$CO on top of CO ice (H$_2$CO/CO)}

\begin{figure*}
  \includegraphics[width=\textwidth]{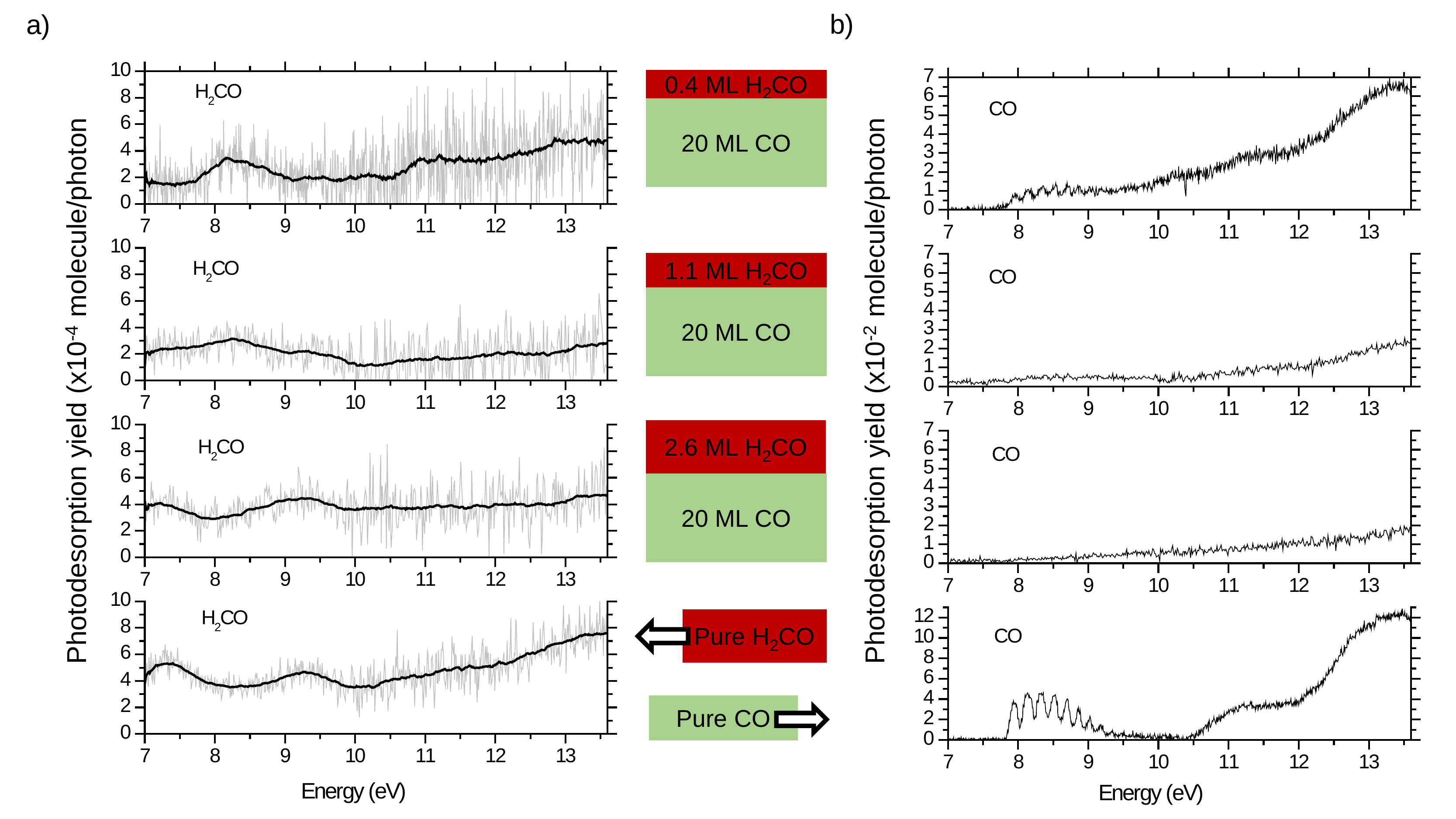}
  \caption{Photodesorption spectra of a) H$_2$CO and b) CO from H$_2$CO/CO layered ices. The different ices are drawn (not to scale) in the center of the figure. The thickness of deposited H$_2$CO is varied from top to bottom: 0.4 ML of H$_2$CO, 1.1 ML and 2.6 ML of H$_2$CO are deposited on top of a CO ice (20 ML). Data on layered ices come from only one energy scan. For H$_2$CO photodesorption (a), the grey line spectra represent raw data, and the black line smoothed data. The bottom layers represent a) H$_2$CO photodesorption from a pure H$_2$CO ice, and b) CO photodesorption from a 20 ML pure CO ice.\citep{dupuy2017a} Note the different vertical scale for the pure CO ice. }
  \label{fig_H2CO_CO}
\end{figure*}

We also studied layered H$_2$CO/CO ices, where H$_2$CO was deposited on top of a CO ice. Photodesorption spectra of H$_2$CO/CO ices, recorded on H$_2$CO (a) and CO (b) masses are presented in Fig.~\ref{fig_H2CO_CO}, for different quantities of H$_2$CO on CO, increasing from 0.4 ML to 2.6 ML, from top to bottom. For comparison, the H$_2$CO spectrum from a pure H$_2$CO ice (like that of Figure~\ref{fig_pure_H2CO}) and the CO spectrum from a pure CO ice \citep{dupuy2017a} are reproduced at the bottom of the figure.

First, let us comment on the photodesorption spectra recorded on the H$_2$CO channel. H$_2$CO photodesorption spectra show different spectral shape and different yields for each ice (Fig.~\ref{fig_H2CO_CO} a), from top to bottom). If only a thin layer of H$_2$CO (0.4 or 1.1 ML) is deposited on top of CO, it seems despite the low signal-to-noise ratio that there is a bump in the 8-9 eV region (see Supporting Information). This could be the signature of CO electronic excitation, transferred to surface H$_2$CO that can desorb through a CO-induced desorption i.e. via an indirect DIET mechanism (see Discussion). Besides, the H$_2$CO signal is not null at 7~eV nor between 9.5 and 10.5 eV, where CO photodesorption is inefficient, indicating that there is a small contribution of H$_2$CO photodesorption from H$_2$CO direct excitation. 
In the intermediate thickness case (middle panel of Fig.~\ref{fig_H2CO_CO}), there are both CO and H$_2$CO contributions, the H$_2$CO contribution being larger than in the 0.4 ML case, and that of CO smaller. When 2.6 ML of H$_2$CO lie on CO, H$_2$CO desorption signal resembles the one from pure H$_2$CO. Thus no strong effect of H$_2$CO thickness on the photodesorption signal is observed, for thicknesses larger than 3 ML. For thicknesses smaller than 3~ML, we see that the H$_2$CO photodesorption yield depends on the ice thickness and on the ice composition (effects of the underlying CO molecules).

 Regarding CO photodesorption spectra, they also change in intensity and in shape with the amount of covering H$_2$CO. When 0.4 ML of H$_2$CO is deposited, CO signatures are clearly seen in the 8--9 eV region, and the spectrum looks rather like that from pure CO ices. It is interesting to see how much CO yields decrease as soon as it is covered with more than 1 ML of H$_2$CO. In this case, CO desorption looks like that from pure H$_2$CO, that is CO comes from photodissociated H$_2$CO and at least does not result from direct excitation of CO. 

Previous studies have shown that photodesorption is a surface process (eg Ref.~\citenum{bertin2012} for a CO ice). All the results on layered H$_2$CO/CO ices in this study also confirm that photodesorption is a surface process, mostly depending on the first 3 monolayers composition, in this case.

\subsection{H$_2$CO and CO mixed ice (H$_2$CO:CO (1:3))}

\begin{figure}
	\includegraphics[width=8.26cm]{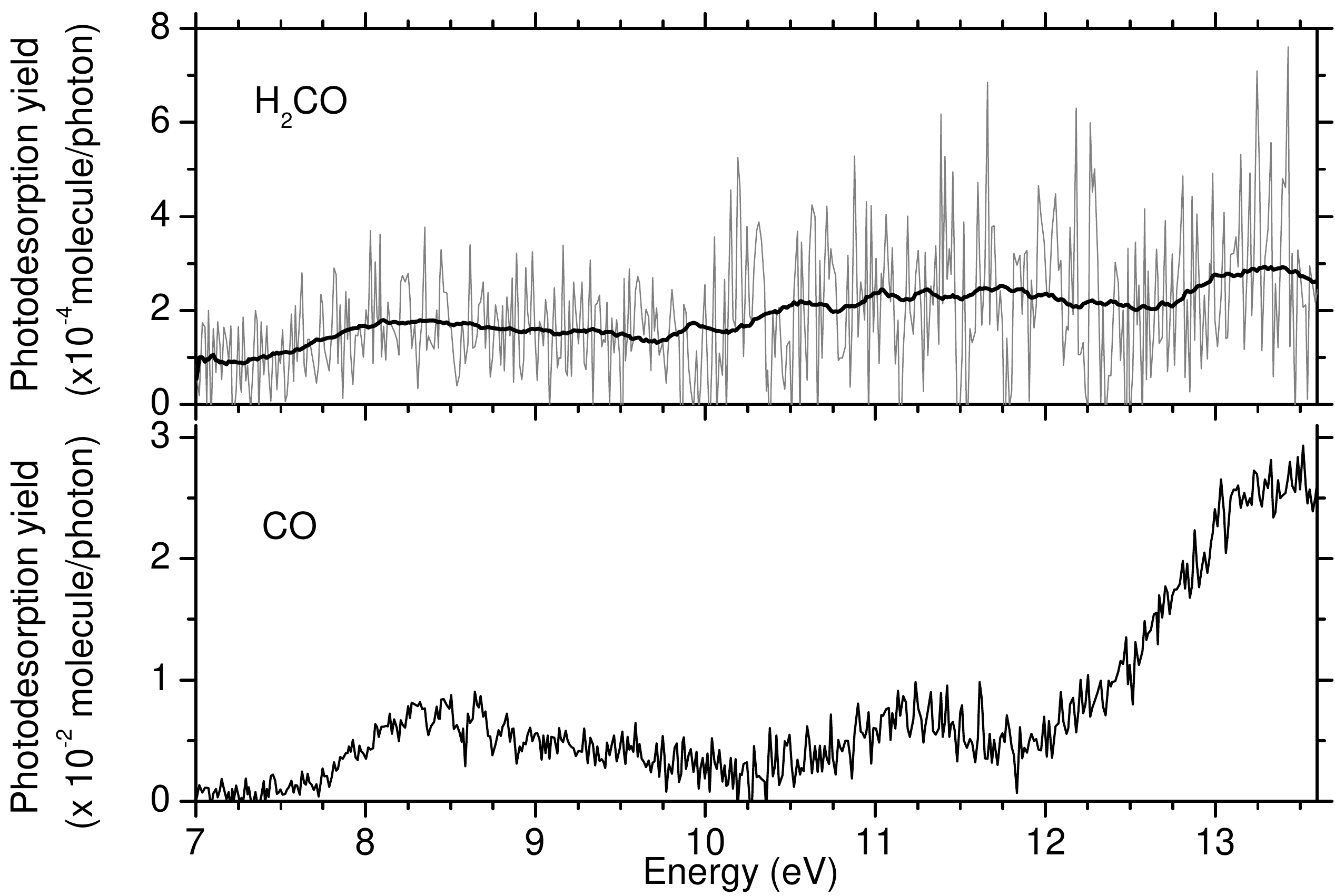}    
       \caption{ Photodesorption spectra of a 24 ML thick H$_2$CO:CO (1:3) mixed ice at 10~K between 7 and 13.6~eV, recorded on  the H$_2$CO mass (m/z 30,\textit{upper panel}) and the CO mass (m/z 28, \textit{lower panel}). Smoothed data are represented by a thick black line in the upper panel.}
     \label{fig_H2CO_CO_mix}
\end{figure}

Photodesorption spectra of a mixed H$_2$CO:CO (1:3) ice are presented in Fig. \ref{fig_H2CO_CO_mix}. Assuming a homogeneous ideal mixing in the solid phase, the surface of this ice is thus composed of $\sim 25\%$ of H$_2$CO and of $\sim 75\%$ of CO. Due to signal/noise limitation, it is not possible to comment on the shape of H$_2$CO photodesorption spectrum. It is however possible to estimate if there is a CO-induced effect in the H$_2$CO photodesorption signal in the mixed ice. H$_2$CO photodesorption efficiencies between 8 and 9~eV are smaller for the mixed ice than for the pure ice, as there is less H$_2$CO available at the surface. After correcting the H$_2$CO photodesorption efficiency by the dilution factor, it is two times larger in the mixed ice than in the pure ice, showing that there is certainly a CO-induced desorption of H$_2$CO in the mixed ice. 

CO photodesorption spectrum from H$_2$CO:CO was also recorded. It looks like that from pure CO, except it is less intense but this is consistent with the fact that there is less CO at the surface of the mixed ice than in the thick pure ice of \citet{dupuy2017a}. As for the layered ice, these results illustrate how photodesorption depends on the ice composition and the thickness. 

\section{Discussion}

\subsection{Photodesorption mechanisms}
\label{section_discussion_psd}

\subsubsection{Mechanisms involved in the photodesorption of H$_2$CO and of CO (pure H$_2$CO ice)}

First of all, we develop some aspects of H$_2$CO photochemistry, that are essential to the discussion of possible mechanisms involved in the photodesorption of H$_2$CO and of CO, for which the excitation of H$_2$CO in dissociative electronic states is certainly the very first step.

Studies of the photolysis of condensed H$_2$CO in the VUV range have shown the existence of these two channels:\cite{thomas1973,glicker1976} \\
H$_2$CO $\xrightarrow{h\nu}$ HCO + H (radical channel) \textit{(i)} \\
H$_2$CO $\xrightarrow{h\nu}$ CO + H$_2$ (molecular channel) \textit{(ii)} \\

As it requires at least 3.5 eV to dissociate H$_2$CO in (i) and (ii) (Refs~\citenum{suto1986,heays2017} and references therein), VUV irradiations provide enough energy to access these channels.

Regarding the present VUV irradiation of solid H$_2$CO, photodesorption, TPD and IR measurements show that CO and H$_2$ are photoproducts present in the ice. This reveals that channel (ii) is a major channel, directly enriching the ice with CO and H$_2$. Besides, the unclear detection of HCO formation in the bulk of our ices could be due to a lack of IR sensitivity or could mean that its destruction pathways are favored over its formation pathways. The HCO radical can be formed from H$_2$CO dissociation through channel (i) or from the reaction between an electronically excited CO (CO*) and H$_2$\cite{chuang2018} (CO* + H$_2$ $\rightarrow$ HCO + H), but it could further evolve to give back H$_2$CO and/or CO (through HCO + H (addition) $\rightarrow$ H$_2$CO which could be barrier-less,\cite{fuchs2009} through HCO + H (abstraction) $\rightarrow$ CO + H$_2$\cite{minissale2016a}, through HCO + HCO  $\rightarrow$ H$_2$CO + CO which is barrier-less,\cite{butscher2017} or through HCO $\xrightarrow{h\nu}$ CO + H). In the end, the final products of the reactions involving HCO are CO and H$_2$CO, the ones that we observe in the bulk in our experimental conditions.  
CO$_2$ was also observed in the ice (see Results), and it could be formed from the following reaction : CO* + CO $\rightarrow$ CO$_2$ + C.\cite{loeffler2005}
It is important to keep in mind that the nature and the amount of photoproducts present in the ice depend on the fluence. For example, CH$_3$OH is produced in the work of \citet{gerakines1996}, for fluences higher than 10$^{17}$ photons/cm$^2$. In our experiment, we did not see any CH$_3$OH band, certainly due to a congestion in the IR spectrum, but that does not preclude its presence, even if certainly weak.

The formation of these CO and H$_2$ photoproducts was also detected in the H$_2$-lamp irradiation of H$_2$CO ices. \citep{gerakines1996,butscher2016} In addition, more complex molecules (glycolaldehyde,  ethylene glycol and the formaldehyde polymer, polyoxymethylene (POM)) were also observed in \citet{butscher2016}'s experiments where a higher fluence than ours was used.

\begin{figure*}
    \includegraphics[width=\textwidth]{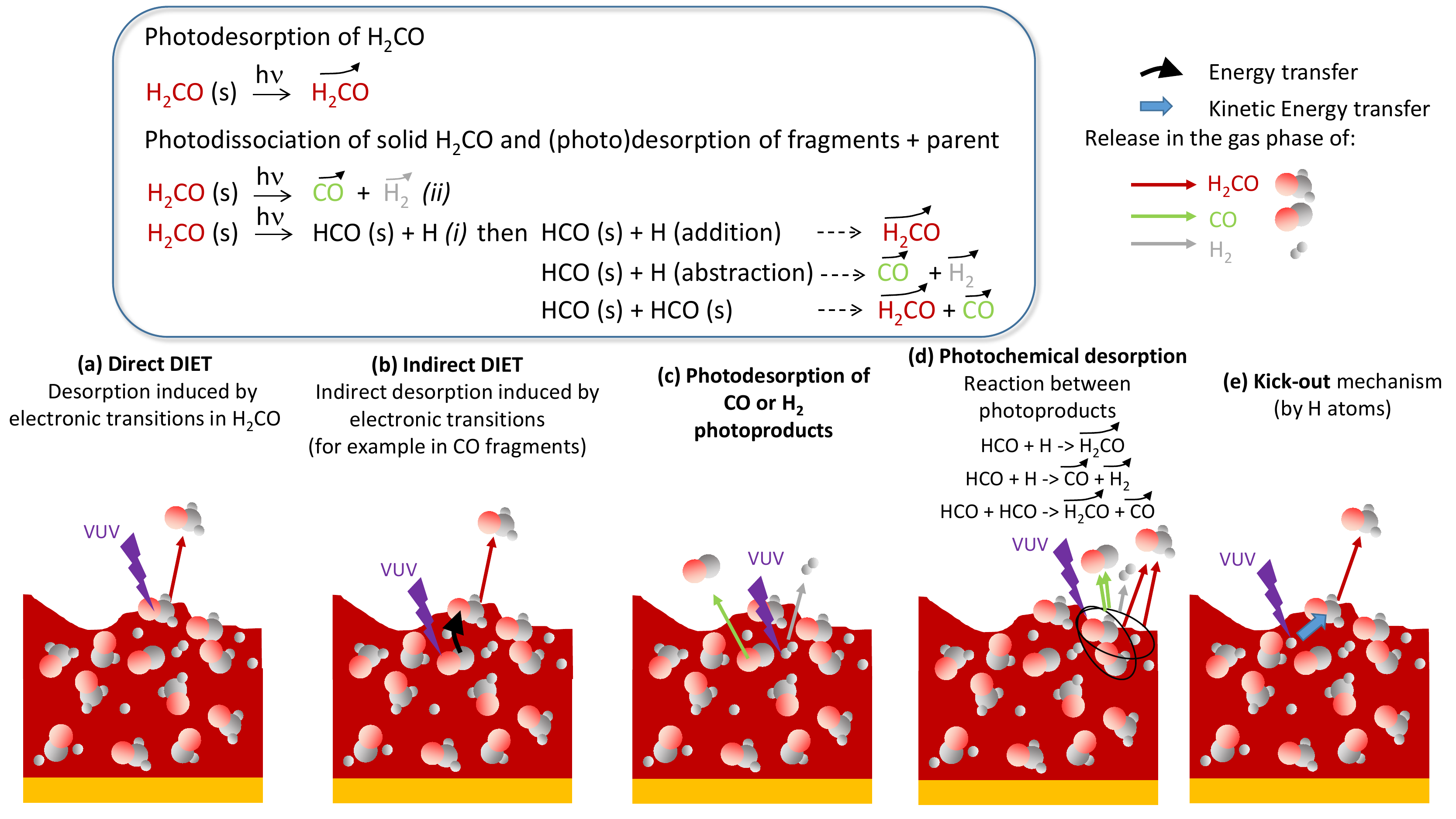}
    \caption{Simplified scheme of the interplay between solid and gas phase following the irradiation of pure H$_2$CO ices. Molecules represented with a curved arrow above them can desorb if they are located near the surface and if they have enough energy. Otherwise, they remain trapped in the bulk of the ice. (s) means that the molecule is located at the surface of the ice, that is, approximately in the first three monolayers.
In the lower part of the figure are also represented five different possible mechanisms that could lead to the photodesorption of H$_2$CO, of CO or of H$_2$ from a pure H$_2$CO ice.     
The release of H$_2$CO in the gas phase includes desorption driven by electronic transitions either in a direct (a) or indirect (b) way, and possibly photochemical desorption (d) and kick-out by H atoms (e). CO photoproducts can desorb right after H$_2$CO photodissociation (c), or can result from the reaction between HCO photoproducts (d), or be induced by the excitation of CO molecules (not shown).}
    
    \label{fig_mechanisms}
\end{figure*}

There is a competition between the reaction of photoproducts and their desorption, if they have enough kinetic energy and if they are located near the surface.  
Based on all the experimental results and on the possible photodissociation paths, the photodesorption of H$_2$CO and of CO fragments could occur through different mechanisms, that are sketched in figure \ref{fig_mechanisms}:

(a) the electronic excitation of an H$_2$CO molecule could lead to its direct desorption (DIET), if it is located at the surface and if it has not photodissociated.

(b) the excitation of a H$_2$CO molecule or of a CO fragment in the first three upper layers of the ice could be transferred to a surface H$_2$CO molecule that could consequently desorb. This process has been named indirect DIET, \cite{bertin2012} and it could contribute to H$_2$CO photodesorption from pure ices, once it is photoprocessed and contains some CO fragments. Indeed, experiments on H$_2$CO/CO layered ices and on H$2$CO:CO mixed ices have shown that CO-induced desorption of H$_2$CO is at play, so that it could also contribute to H$_2$CO photodesorption efficiency from pure H$_2$CO ices, if enough CO photofragments are present in the ice. As approximately 20\% of the ice is CO at a fluence of $3 \times 10^{16}$ photons cm$^{-2}$, this mechanism is possible. 

(c) the photodesorption of CO and H$_2$ products was also observed and it is an important finding. It could happen for CO and H$_2$ if they are located near the surface and if they have enough energy right after H$_2$CO dissociation, or (d) result from surface reaction processes such as reactions between HCO photoproducts\cite{butscher2017} or H atom abstraction of an HCO photoproduct,\cite{minissale2016a} or be induced by the excitation of CO molecules, as the electronic excitation of CO is probably seen in Fig.~\ref{fig_pure_H2CO_aged}.

(d) photochemical desorption of H$_2$CO could occur, through recombination of HCO + H photofragments (HCO directly coming from H$_2$CO dissociation, or from CO*+H$_2$). Indeed the reaction is exothermic (by 3.9 eV), so that it releases an energy much larger than the H$_2$CO binding energy (0.324 eV),\cite{noble2012} and it could possibly lead to a photodesorption event if it occurs near the surface. Recombination of two HCO radicals could also produce H$_2$CO\cite{butscher2017} that could be released in the gas phase, together with CO fragments.  

(e) an H atom could kick-out a H$_2$CO molecule present at the surface. Such a kick-out mechanism by H atoms has been predicted for H$_2$O.\citep{andersson2008} It is expected that the efficiency of this mechanism decreases with the mass of the target molecule, so as the mass of H$_2$CO is larger than the mass of H$_2$O, this process will be less favorable for H$_2$CO ice than for H$_2$O ice.

Therefore this set of data shows that desorption driven by electronic transitions either in a direct or indirect way contribute to the photodesorption of H$_2$CO, possibly together with photochemical desorption. 
For CO$_2$ ice, the photochemical desorption mechanism, through recombination of CO and O, was quantitatively measured to be 10\% of the total photodesorption efficiency.\cite{fillion2014} On the contrary, for CH$_3$OH, the photochemical recombination CH$_3$O + H is proposed as a possible desorption pathway.\cite{bertin2016} Through the present study of the photodesorption of H$_2$CO, we have hints of the presence of photochemical recombination through the dissociative nature of electronic states and the presence of photoproducts. However it does not allow one to estimate the relative contribution of photochemical recombination with respect to DIET or indirect DIET mechanisms, all these mechanisms being certainly at play.

H$_2$CO photodesorption efficiency is larger than CH$_3$OH:\citep{bertin2016,cruz-diaz2016} H$_2$CO photodesorption yields from pure H$_2$CO ices are $\sim 30-40$ times higher than CH$_3$OH from pure CH$_3$OH ices (Table~\ref{table_yields}). When discussing photodesorption efficiencies from one molecule to the other, it is convenient to compare yields in molecule per absorbed photon, and not in molecule per incident photon (as derived here). For H$_2$CO, it is not possible to obtain yields in molecule per absorbed photon, as the absorption cross section in the solid phase is missing. As absorption cross sections of different ices differ only by a factor 2 or 3,\cite{cruz-diaz2014b} the large photodesorption efficiency of H$_2$CO cannot be explained by differences in absorption cross sections and another parameter has to be invoked. It cannot be the mass of H$_2$CO and of CH$_3$OH, as they are very similar, but it could be differences in binding energies (larger for CH$_3$OH than for H$_2$CO), in vibrational degrees of freedom (12 for CH$_3$OH versus 6 for H$_2$CO), or in the fragmentation of these two molecules.

In addition to the photodesorption of intact H$_2$CO, the fact that CO photofragments desorb is an important finding, and we proposed several mechanisms leading to its photodesorption. It seems that in our experiments, CO desorption decreases with fluence (compare Fig.~\ref{fig_pure_H2CO} and Fig.~\ref{fig_pure_H2CO_aged}), that could be explained by the consumption of CO to give HCO and H$_2$CO products, or by the modification of the ice surface, so that in the end less molecular fragments photodesorb. 
While the deposited H$_2$CO ice is amorphous,\cite{hiraoka2005} modifications of the ice structure can happen due to VUV irradiation. Possible modifications include an amorphisation or a compactification of the H$_2$CO ice (the latter is observed in the irradiation of water ice).\cite{palumbo2010,dartois2018}

\subsubsection{CO-induced desorption of H$_2$CO (H$_2$CO/CO layered ices and H$_2$CO:CO mixed ices)}

To unravel the efficiency of CO-induced desorption, several layered ices have been studied.\cite{bertin2012,bertin2013,dupuy2017a} The characteristic signatures of CO vibronic excitation were observed in the photodesorption pattern of the molecule X above CO. This indicates a transfer of energy from CO molecules to the surface molecules, the indirect DIET process.\citep{bertin2012} The maximum photodesorption yields of X species in the 8--9 eV region for X/CO ices are : $2.5 \times 10^{-2}$ molecule/photon for X=N$_2$,\citep{bertin2013} $2 \times 10^{-3}$ molecule/photon for X=CH$_4$,\citep{dupuy2017a} and $3 \times 10^{-4}$ molecule/photon for X=H$_2$CO (this work). These yields vary for the different X molecules. The efficiency of the indirect DIET mechanism depends on several parameters: the inter-molecular energy transfer, the intra or inter-molecular energy relaxation once it has been transferred, and the binding energy of the molecule. 

The comparison between N$_2$ and CH$_4$ yielded interesting findings and pointed out the important role of intra- or inter-molecular energy relaxation in this system (the binding energy of the two molecules on CO should be approximately the same).\citep{dupuy2017a} 
Comparison between N$_2$ and H$_2$CO indirect desorption also gives interesting conclusions. The mass of N$_2$ and H$_2$CO is approximately the same, so a simple kinetic momentum transfer between CO and N$_2$ or H$_2$CO cannot explain the different efficiencies. The binding energy of H$_2$CO on CO (not known) is expected to be larger than the binding energy of N$_2$ on CO (because H$_2$CO has a permanent dipole moment whereas N$_2$ has none). A higher binding energy of H$_2$CO on CO could thus play a role in the quenching of the CO-induced efficiency. We also expect H$_2$CO to relax the vibrational excess energy more efficiently than N$_2$, because of the degrees of freedom in H$_2$CO with respect to diatomics. However, the vibrational degrees of freedom of H$_2$CO (6) are smaller than for CH$_4$ (9), whereas H$_2$CO desorbs less efficiently than CH$_4$ through indirect DIET. All these results on different systems show that both energy relaxation and large binding energies play a role on quenching the CO-induced photodesorption.

Mixed ices containing CO and N$_2$ or CH$_3$OH were also studied in \citet{bertin2013,bertin2016}. In the CO:N$_2$ binary ice,\citep{bertin2013} the photodesorption spectrum is a linear combination of that of CO and N$_2$. In the H$_2$CO:CO mix (Fig. \ref{fig_H2CO_CO_mix}) or in the CH$_3$OH:CO mix,\citep{bertin2016} this is not the case, as the H$_2$CO or CH$_3$OH photodesorption spectrum is influenced by photochemical processes and surface modification. As already shown by H$_2$CO/CO experiments, CO-induced desorption of H$_2$CO is less efficient than N$_2$ indirect desorption. This shows once more that photodesorption depends on the ice composition and on the considered molecule, the main difference between N$_2$ on one side and H$_2$CO and CH$_3$OH being that the last two photodissociate and have a large binding energy.

\subsection{Photodesorption yields in various astrophysical media}

\label{discussion_table}

\begin{table*}
      \caption[]{Average photodesorption yields $Y_i$ ($\times$ 10$^{-4}$ molecule per incident photon; $i$ is H$_2$CO or CO) for pure H$_2$CO ice and a mixed H$_2$CO:CO ice, in various interstellar environments with radiation fields from 7 to 13.6~eV. Average photodesorption yields of CH$_3$OH from pure CH$_3$OH ice are also reported.\citep{bertin2016}}

         \begin{tabular}{c c c c c c c}
            \hline
 Ice   		 &  Photodesorbed 	&	ISRF\textsuperscript{\emph{a}} 	&	PDR\textsuperscript{\emph{b}}  	& PDR\textsuperscript{\emph{b}}  		& Secondary UV \textsuperscript{\emph{c}} & Protoplanetary disk\\
     		&		species	$i$	&				&    A$_V$=1		&     A$_V$=5  &       			& TW Hya\textsuperscript{\emph{d}}     \\    
              									& \multicolumn{6}{c}{Y$_i$($\times$ 10$^{-4}$ molecule/photon)} \\
            \hline
          
Pure H$_2$CO 	&		H$_2$CO		&   	5		&	4			&		4	&		4		&		4		\\	
			&		CO			&		45		&	28			&		22	&		45			&		40			\\
\hline

H$_2$CO:CO (1:3)&		H$_2$CO 		&		8\textsuperscript{\emph{$\dagger$}}		&		6\textsuperscript{\emph{$\dagger$}}			&	6\textsuperscript{\emph{$\dagger$}}			&		6\textsuperscript{\emph{$\dagger$}}				&		10\textsuperscript{\emph{$\dagger$}}				\\
			&		CO			&		61		&		40		&	38		&		42			&		26			\\
            \hline
Pure CH$_3$OH\textsuperscript{\emph{e}} &  CH$_3$OH  		&   0.12   		&				&			&		0.15		&					\\
	 			&  H$_2$CO  		&     0.07 		&				&			&		0.12		&					\\
         \end{tabular}
  
  UV fields between 7 and 13.6 eV are taken from :        
  \textsuperscript{\emph{a}} \citet{mathis1983};
  \textsuperscript{\emph{b}} PDR Meudon Code \citet{lePetit2006} (see text for details);
  \textsuperscript{\emph{c}} \citet{gredel1987};
  \textsuperscript{\emph{d}} \citet{heays2017} (FUV observation of TW-Hydra from \citet{france2014}, extrapolated to a broader spectral range);
  \textsuperscript{\emph{e}} \citet{bertin2016}
 
  \textsuperscript{\emph{$\dagger$}}	This average photodesorption yield is normalized to the fraction of the surface of the ice containing H$_2$CO $f_s$ (see text for details)

         \label{table_yields}
   \end{table*}

Average photodesorption yields $Y_i$ are reported in Table~\ref{table_yields}, for $i=$ H$_2$CO and CO, from the pure H$_2$CO ice and from the mixed H$_2$CO:CO ice, for different UV radiation fields.
They are derived from experimental photodesorption spectra (Fig.~\ref{fig_pure_H2CO}) and radiation fields in different interstellar environments, as described in Ref.~\citenum{dupuy2017}. 
In the particular case of the mixed H$_2$CO:CO ice, average photodesorption yields $Y_{H_2CO}$ are obtained through the following :

$Y_{H_2CO} = \frac{Y_{H_2CO}^{measured}}{f_s},$

with $f_s$ the fraction of the surface of the ice containing H$_2$CO, which is 0.25 for the H$_2$CO:CO (1:3) mixed ice. These photodesorption yields are given in molecule per incident photon, so that they can be directly implemented in astrochemical models, wihout any correction by the number of monolayers included in the surface. For any molecule in the upper three monolayer, photodesorption yields of Table~\ref{table_yields} can be added to models.
  
Average H$_2$CO photodesorption yields are in the same range for the different regions explored : the Interstellar Radiation Field, PDR radiation fields, secondary UV photons or a typical PPD spectrum. This is due to the shape of the photodesorption spectrum (Fig. \ref{fig_pure_H2CO}), which do not vary significantly with the photon energy. 
However, there is an effect of the ice composition, as H$_2$CO photodesorption yields vary if pure or mixed ices are considered (Table~\ref{table_yields}). They are slightly larger when H$_2$CO is mixed with CO, which is an enhancement certainly due to the CO-induced photodesorption of H$_2$CO. Thus to model an ice whose surface contains H$_2$CO and CO, it is more appropriate to use the average photodesorption yield from the CO-containing ice than from the pure ice.

Photodesorption yields of Table \ref{table_yields} are valid for H$_2$CO present at the surface (that is in the first 3 ML) of an ice whose total thickness is larger than 3 ML. If one considers astrophysical environments or time periods where ices start to form on interstellar grains, only thin ices ($\lesssim 1$ ML) cover the grain substrate, and photodesorption yields could be different. Indeed, the underlying substrate could change the photodesorption, as a function of wavelength and in intensity, depending on its nature. If we consider Fe-silicates, they absorb in the VUV range (see eg Ref. \citenum{kohler2014}), so an energy transfer to the thin ice above could be possible and change the photodesorption yields. 
Besides, for these thin ices, the presence of pores and of inhomogeneities of the grain substrate can change the adsorption of molecules and their arrangement, which could result in different photodesorption mechanisms and efficiencies. 
This was observed for CO deposited on porous-amorphous H$_2$O, where  pores in the H$_2$O ice reduce the CO photodesorption yield.\cite{bertin2012} 
It is thus probable that the nature of the surface of interstellar grains and their morphology affect photodesorption if thin H$_2$CO-containing layers of ices are involved, and laboratory experiments are needed to characterize and quantify this.

It is possible to compare the H$_2$CO photodesorption yields from pure ices with H$_2$CO photodesorption yields from previously studied ices. H$_2$CO was also observed as a photodesorbing fragment from CH$_3$OH ice.\citep{bertin2016,cruz-diaz2016} However, yields are around $10^{-5}$ molecule/photon (values from \citet{bertin2016} are reproduced in Table \ref{table_yields}), which is much smaller than H$_2$CO from pure H$_2$CO ices, $ > 4 \times 10^{-4}$ molecule/photon (Fig.~\ref{fig_pure_H2CO_aged} and Table \ref{table_yields}) i.e., H$_2$CO photodesorption from pure H$_2$CO ices is 50 times more efficient than from pure CH$_3$OH ices. In addition to CH$_3$OH ices,  H$_2$CO desorption has been detected from ethanol or H$_2$O:CH$_4$ ice.\citep{martin-domenech2016} In this experiment, H$_2$CO photodesorption yields of $6 \times 10^{-4}$ molecule/photon and $4.4 \times 10^{-4}$ molecule/photon were found, respectively. These are the same values as found in the present study, meaning that H$_2$CO photodesorption has to be taken into account both from H$_2$CO and from the recombination of photofragments at the surface of interstellar ices. In the end, all desorption channels have to be taken into account, when possible.\cite{guzman2013}

If we consider pure H$_2$CO or pure CH$_3$OH ices, CO fragment is the major species that desorbs. CO desorption from H$_2$CO ices is $4.5 \times 10^{-3}$ molecule/photon (Table~\ref{table_yields}).
This yield is directly derived from our measurements, so that it can be added in models, without any photodissociation ratio correction. 
We also found that CO photodesorption from H$_2$CO ices is only 2 times less than CO desorption from pure CO ices .\citep{fayolle2011} CO ice abundance (20\% relative to H$_2$O ice, cf Ref.~\citenum{oberg2016}) is larger than H$_2$CO ice abundance (a few percent relative to H$_2$O ice), but CO photodesorption from H$_2$CO and more generally from organic molecules could be taken into account in models.

Pure H$_2$CO ices constitute model ices, however astrophysical ices are much more complex, and H$_2$CO could be mixed with CO, on a CO ice, below CO ice, or with H$_2$O or CO$_2$. The mixture of H$_2$CO in a H$_2$O-dominated ice could for example give different photodesorption yields and mechanisms, because of the dangling O-H bonds of surface H$_2$O who rapidly and efficiently evacuate excess vibrational energy\cite{Zhang2011}. Besides, the phase of interstellar H$_2$O, if porous/compact amorphous\cite{dartois2015a} or crystalline,\cite{mcclure2015} could also influence photodesorption. Precisely taking into account properly the effects of the ice composition on the photodesorption of H$_2$CO necessitates a dedicated study.  Still, from the H$_2$CO:CO mixed and H$_2$CO/CO layered ices studied here, we see that the abundance of H$_2$CO in the first three monolayers has to be taken into account, together with CO-induced desorption effects.
Photodesorption yields adapted to the modelled interstellar ice have to be considered.

\subsubsection{Prestellar cores and PDR}

H$_2$CO in the gas phase has been observed in dense cores and in low or high-UV flux PDR (e.g. Refs \citenum{bacmann2003,guzman2011,guzman2013,cuadrado2017}), and UV photodesorption has often been called upon to explain the observations.

In dense cores, H$_2$CO in the gas phase could come from different processes : it could be formed directly in the gas, or desorb from grains through chemical desorption, cosmic-ray sputtering, or UV photodesorption. 
Chemical desorption is often considered in astrochemical models,
\cite{garrod2006} however little experimental data exist, so that the efficiency taken in models is highly uncertain, especially when considering reactions occurring on icy grains. Indeed, experimental results of \citet{minissale2016b,minissale2016a} have shown that the chemical desorption process is much less efficient on icy grains than on bare grains, and sometimes too weak to be detected.   
Whereas H$_2$CO chemical desorption from \textit{bare} grains has important effects on gas phase abundances, those changes were much smaller when considering chemical desorption from water \textit{icy} grains.\citep{cazaux2016} 
The competition between UV photodesorption efficiency of H$_2$CO and chemical desorption efficiency on icy grains can now be modelled.

During the formation of interstellar ices in the early phases of core contraction, UV photodesorption plays an important role, as shown by \citet{kalvans2015}. At this early phase, the ISRF can still penetrate the cloud, so that photodesorption mostly governs the onset of ice accumulation onto grains.
Later on, whereas the ISRF is shielded, there are nonetheless UV photons emitted by H$_2$ excited by cosmic rays (called secondary UV photons). H$_2$CO average photodesorption yields from secondary UV photons in the dense cores are the same as from the ISRF (Table~\ref{table_yields}). The secondary UV flux in dense cores is 10$^4$ photons cm$^{-2}$ s$^{-1}$.\cite{Shen2004} If we consider the beginning of H$_2$CO ice formation at 1000 years\cite{cuppen2009}, and a dense cloud lifetime of 10$^6$ years, this gives a fluence of $3 \times 10^{14}$ to $3 \times 10^{17}$ photons cm$^{-2}$. In our experiments the fluence was at most $3 \times 10^{16}$ photons cm$^{-2}$, which is thus representative of that in dense clouds, except in the last 10$^5$-10$^6$ years of the core. These yields could thus be implemented in models, and their effects on dense cores at various time explored.  

In low-UV illuminated PDR, nonthermal desorption from H$_2$CO-containing icy grains is necessary to reproduce the observed abundances,\citep{guzman2011,guzman2013} and photodesorption was considered to be a very good candidate. 
As the effects of VUV radiation on the photodesorption of H$_2$CO ices were not known, these authors considered two different pathways: the first one is the release of H$_2$CO in the gas phase with a yield of $5 \times 10^{-4}$ molecule per photon, and the second one is the photodissociation and release of HCO and H fragments, with a yield of $5 \times 10^{-4}$ molecule per photon. 
Our experimental work reveals that different yields and pathways have to be considered (Table~\ref{table_yields}): the H$_2$CO photodesorption yield is $4-6 \times 10^{-4}$ molecule per photon, the HCO + H path was not observed; whereas the CO + H$_2$ was the most efficient path giving $22-28 \times 10^{-4}$ CO molecule per photon. Taking these findings into account could result in different gas phase abundances of H$_2$CO and related molecules.

In order to study the effect of the radiation field on the photodesorption efficiency, radiation fields at different extinctions were generated with the Meudon PDR code,\cite{lePetit2006} at A$_V$=1 and A$_V$=5. Low-flux PDR conditions corresponding to the Horsehead PDR were considered, with an incident field of 80 times that of \citet{mathis1983}, and a density profile as described in \citet{guzman2011}. A$_V$=1 is the closest to the star, whereas A$_V$=5 is further in the PDR. We assume that H$_2$CO ice could start to form at A$_V$=1, like CH$_3$OH ice.\cite{esplugues2016} 
 When going further in the PDR, the high-energy part of the spectrum is strongly attenuated. Despite the spectral differences at A$_V$=1 and A$_V$=5, average photodesorption yields are approximately the same from an extinction to the other mainly because photodesorption spectra do not vary significantly with the photon energy. As a consequence, for H$_2$CO, a single photodesorption yield can be used at all extinctions. H$_2$CO photodesorption rates should thus match the diminishing photon flux in the cloud.  
One should keep in mind that in PDR, UV photon fluxes vary strongly: if we consider a flux of $2 \times 10^{10}$ photons cm$^{-2}$ s$^{-1}$ at the edge of the PDR, it will decrease down to $2 \times 10^{2}$ photons cm$^{-2}$ s$^{-1}$ at A$_V$=5 (considering only dust extinction). The fluence over the whole PDR lifetime ($5 \times 10^5$ years)\cite{Pound2003} thus varies from $3 \times 10^{23}$ photons cm$^{-2}$ at the edge of the PDR to $3 \times 10^{15}$ photons cm$^{-2}$ at A$_V$=5 and at 12~eV. As a consequence, ices in PDR regions experience various irradiation conditions, from those similar to the present study, to others where photoproducts could appear and the ice composition change, possibly affecting the photodesorption yield of H$_2$CO and of CO, but also introducing other photodesorbing species.

The Orion bar case is different from the Horsehead PDR. It is a high-UV flux PDR, so dust temperatures are around 55--70~K, which is higher than in the Horsehead PDR (around 20-30 K, \citet{guzman2011}). So there is a competition between thermal and nonthermal desorption, contrary to the Horsehead case. But keep in mind that the adsorption energy of pure H$_2$CO is 3770~K and the adsorption energy of H$_2$CO bound to H$_2$O is 3259~K,\citep{noble2012} the latter being much larger than the one usually considered from \citet{garrod2006b}, 2050 K. Implementing an accurate adsorption energy is necessary as variations in adsorption energies induce variations in ice abundances and ice composition (see eg \citet{penteado2017}). This could mean that H$_2$CO thermal desorption is overestimated in some models and in some environments such as high UV flux PDR, and that non-thermal desorption could play a role. We can estimate that below $\sim$ 76~K\footnote{For this estimation, we equalled the photodesorption rate with the thermal desorption rate. For UV photodesorption, we considered a UV flux of $10^{10}$ photons cm$^{-2}$ s$^{-1}$. For thermal desorption, we used a first order kinetics with a prefactor of $10^{28}$ mol. cm$^{-2}$ s$^{-1}$ and an adsorption energy of 3765~K from \citet{noble2012}.}, UV photodesorption dominates over thermal desorption. As dust temperatures are lower than 76~K in the Orion Bar, UV photodesorption is thought to impact the gas-grain ratio in this high-flux region.   
This was actually (indirectly) explored in the work of \citet{esplugues2016,esplugues2017}. As soon as CO and H$_2$CO photodesorptions are taken into account in high-flux PDR, there are changes of several orders of magnitude in the gas phase abundances of CO and H$_2$CO, so that UV photodesorption clearly plays a role.

\subsubsection{Protoplanetary disks (PPDs)}

Several spatially-resolved observations of H$_2$CO in the gas phase have been performed on disks around a T Tauri (eg Ref.~\citenum{oberg2017}) or a Herbig star.\cite{carney2017}
These observations, together with models, showed that there are several H$_2$CO components in these disks. Whereas close to the star, warm H$_2$CO should form in the gas phase, further from the star, grain surface formation (which is active in extended areas of disks \cite{aikawa1999,walsh2014}) and non-thermal desorption are both needed to explain the observations.
More precisely, the detection of gas phase H$_2$CO in the outer parts of the disk, where CO is frozen on grains, favors the role of surface processes and of nonthermal processes such as photodesorption by UV or X-ray photons.\cite{loomis2015,oberg2017,carney2017,podio2019}
Complex models of PPDs have demonstrated that in the midplane, non-thermal desorption processes such as cosmic-ray-induced thermal desorption, X-ray spot heating and photodesorption by internal and external UV photons are necessary to produce H$_2$CO in the gas phase.\cite{walsh2014} Now that experimental data on the photodesorption of H$_2$CO ices are available, a precise study could be performed, with the determination of the relative efficiency of each non-thermal process.   

In protoplanetary disks such as TW Hydra, the UV flux varies from 10$^3$ to 10$^9$ photons cm$^{-2}$ s$^{-1}$ at different locations in the disk, approximately where ices exist. With time spanning from 1 to 10 million years, this gives fluences of $3 \times 10^{16} - 3 \times 10^{17}$ photons cm$^{-2}$ for the areas which receive less photons, to $3 \times 10^{22} - 3 \times 10^{23}$ photons cm$^{-2}$ for the areas which are strongly illuminated. Our experimental fluence reproduces the low-illuminated regions of disks, whereas in the strongly-illuminated regions, photodesorption yields could differ if the ice composition is strongly altered.

As already mentioned, H$_2$CO is connected with CH$_3$OH, which has been recently detected in PPDs.\citep{walsh2016} Given all the experimental and modelling work performed on CH$_3$OH, discussing some findings on CH$_3$OH photodesorption can directly inspire what could be done with H$_2$CO: 

(i) Varying the CH$_3$OH photodesorption yield \citep{oberg2009b,bertin2016,cruz-diaz2016} has drastic consequences 
on the abundance of both solid and gas phase CH$_3$OH by orders of magnitude, in some parts of the disk. \citep{walsh2017,ligterink2018} It should be noted that the experimental CH$_3$OH photodesorption yield of $10^{-5}$ molecule/photon \citep{bertin2016,cruz-diaz2016} is low, but it is \textit{not negligible} as it is high enough to have an effect on the gas replenishment.\citep{walsh2017,ligterink2018}

(ii) The comparison between UV photodesorption and chemical desorption of CH$_3$OH showed that photodesorption largely dominates chemical desorption in the majority of the disk.\citep{ligterink2018}

(iii) Photodesorption do shape snowlines differently than chemical desorption, as it gives CH$_3$OH snowlines deeper in the disk. Therefore photodesorption is important in setting the location of snowlines, especially for non-volatile species such as H$_2$O, CH$_3$OH \citep{ligterink2018,agundez2018} and probably H$_2$CO (the H$_2$CO snowline is located closer to the star than the CO snowline,\cite{qi2013} as its formation on grains only requires that the CO molecule spends some time on the surface of grains \citep{aikawa1999,walsh2014} and as its binding energy is larger than the CO one).

(iv) Interestingly, taking into account the desorption of CO fragments from CH$_3$OH ice has diverse impacts: it changes the abundance of gas phase CO in some parts of the disk
and the CH$_3$OH snowline location.\citep{walsh2017,ligterink2018} Indeed, when no fragmentation is considered, CH$_3$OH re-adsorption competes with photodesorption, whereas when CH$_3$OH photofragmentation is added, CH$_3$OH has to reform on the grains, and that changes the location of the snowline.

(v) Gas phase CH$_3$OH shows a spatial coincidence with H$_2$CO \cite{walsh2016,oberg2017} in the T Tauri disk TW-Hydra. For this T Tauri, the methanol-to-formaldehyde ratio is 1.27,\cite{walsh2016} but it is much smaller (0.24) for a PPD around a Herbig star.\cite{carney2019} One possible explanation for these unequal ratios is that the photodesorption of CH$_3$OH and H$_2$CO is different in these two disks, as the spectral distribution and intensity of the stellar radiation differ between T Tauri and Herbig stars.\cite{carney2019} With wavelength-resolved photodesorption spectra now available, detailed chemical modelling of these two disks can be performed, and the effects of photodesorption characterized.

In the light of all the experimental, observational and modelling work that has been done on CH$_3$OH, experimental data on H$_2$CO ice could be implemented or updated. The implementation of H$_2$CO photodesorption yields and of the branching ratio, giving CO fragments, both larger than in the CH$_3$OH ice case (Table~\ref{table_yields} and Ref.~\citenum{bertin2016}) could give useful information: 
(i) on gas/solid abundances of H$_2$CO and on the gas abundance of CO; 
(ii) on the solid phase abundance of H$_2$CO as it was shown that it drops drastically when UV photodesorption is included in the upper layers of the disks;\citep{walsh2014} besides, any change in the abundance of the 'building block' H$_2$CO will certainly have consequences on other solid phase abundances, such as that of CH$_3$OH and of COMs
(iii) on the H$_2$CO snowline location, as photodesorption is important for the location of snowlines especially in the outer disk for species with large binding energies.

\section{Conclusions}

A quantitative study of the photodesorption from pure and H$_2$CO-containing ice was performed. H$_2$CO photodesorption spectrum presents dissociative electronic states and the photodesorption of CO and H$_2$ fragments was also measured. Photodesorption mechanisms were constrained, including (indirect) desorption induced by electronic transitions and photochemical desorption. 

From the energy-resolved photodesorption yields, we could derive H$_2$CO and CO average photodesorption yields in several astrophysical environments such as the ISRF, PDR, dense cloud or PPD, and found that these yields slightly vary from one environment to the other. These yields can be directly added to astrochemical models, without any branching ratio correction. 

 This study also confirms that photodesorption yields strongly differ from one molecule to the other, for example from CO to H$_2$CO and to CH$_3$OH and also that they depend on the ice composition. Laboratory experiments on ice analogs are a necessary step, as it is not possible yet to predict photodesorption efficiencies.
Elements are gathered to investigate further the H$_2$CO non-thermal desorption in dense cores, PDR and disks, to tune more finely gas phase and solid phase abundances, and possibly snowline locations. Its impact on the gas-to-ice balance has to continue to be explored, for H$_2$CO and all related species, spanning from CO to COMs.

\begin{acknowledgement}

The authors thank SOLEIL for provision of synchrotron radiation
facilities under the project 20150760 and also Laurent Nahon and the DESIRS beamline for their help. This
work was supported by the Programme National 'Physique et Chimie du Milieu Interstellaire' (PCMI) of CNRS/INSU with INC/INP co-funded by CEA and CNES. Financial support from LabEx MiChem, part of the French state funds managed by the ANR within the investissements d'avenir programme under reference ANR-11-10EX-0004-02, and by the Ile-de-France region DIM ACAV programme, is gratefully acknowledged.

\end{acknowledgement}

\begin{suppinfo}

Available free of charge on the ACS Publications website at DOI: 10.1021/acsearthspacechem.9b00057

Flux measurement correction \\
Ageing effects \\
Extrapolation of the photodesorption spectra to lower energies \\
Illustration of the CO contribution in the H$_2$CO/CO layered ice \\

\end{suppinfo}

\bibliography{H2CO}

\end{document}